\documentclass[prd,twocolumn,tightenlines,superscriptaddress,nofootinbib,
showpacs]{revtex4}
\usepackage{amssymb,latexsym}
\usepackage{amsmath,amsbsy,bbm}
\usepackage{epsfig,bm}
\usepackage{graphicx,comment}
\usepackage{color}
\usepackage{soul}
\usepackage{csquotes}
\usepackage[normalem]{ulem}
\begin{document}

\title{Ground state energy density, susceptibility, and Wilson ratio of a two-dimensional disordered quantum 
spin system}

\author{J.-H. Peng}
\affiliation{Department of Physics, National Taiwan Normal University,
88, Sec.4, Ting-Chou Rd., Taipei 116, Taiwan}
\author{D.-R. Tan}
\affiliation{Department of Physics, National Taiwan Normal University,
88, Sec.4, Ting-Chou Rd., Taipei 116, Taiwan}
\author{F.-J. Jiang}
\email[]{fjjiang@ntnu.edu.tw}
\affiliation{Department of Physics, National Taiwan Normal University,
88, Sec.4, Ting-Chou Rd., Taipei 116, Taiwan}

\begin{abstract}
  
  A two-dimensional (2D) spin-1/2 antiferromagnetic Heisenberg model with a
  specific kind of quenched disorder is investigated, using the first principles
  nonperturbative quantum Monte Carlo calculations (QMC). The employed
  disorder distribution has a tunable parameter $p$ which can be considered
  as a measure of the corresponding randomness. In particular, when $p=0$,
  the disordered system becomes the clean one. Through a large scale QMC,
  the dynamic critical exponents $z$, the ground state energy densities $E_0$,
  as well as the Wilson ratios $W$ of various $p$ are determined with high
  precision. Interestingly, we find that the $p$ dependence of $z$ and $W$
  are likely to be complementary to each other. For instance, while the $z$
  of $0.4 \le p \le 0.9$ match well among themselves and are statistically
  different from $z=1$ which corresponds to the clean system, the $W$ for
  $p < 0.7$ are in reasonable good agreement with that of $p=0$.
  The technical subtlety of calculating these physical quantities for a
  disordered system is demonstrated as well. The results
  presented here are not only interesting from a theoretical perspective,
  but also can serve as benchmarks for future related studies.

\end{abstract}

\vskip-0.25cm

\maketitle

\section{Introduction}

Spatial dimension two is extraordinary from a theoretical point of view.
This is because
according to the famous Mermin-Wagner theorem, for finite systems with
short-range interactions, continuous symmetries cannot be
broken spontaneously at any temperature $T>0$ \cite{Mer66,Hoh67,Col73,Gel01,Car96,Sac11}. As a result, for
two-dimensional (2D) quantum spin antiferromagnets (AF), the associated
studies have been focusing on certain exotic finite temperature
properties of the systems.
Particularly, several universal
quantities are predicted and verified numerically. Such a temperature
region where these unusual universal features exist
is called the quantum critical regime (QCR) in the literature and
has been explored in detail during the last few decades.
\cite{Cha89,Chu93,Chu931,Chu94,San95,Tro96,Tro97,Tro98,Kim99,Kim00,San11,Sen15,Tan181}.

For 2D quantum spin AF system, whenever QCR is mentioned, it typically
refers to a finite temperature region. However, such an exotic regime
extends to zero
temperature at a quantum critical point (QCP). In addition, the (finite $T$)
region above a QCP is where these profound
characteristics can be uncovered the most clearly \cite{Tro98,Tan181}.

A physical observable, namely the spinwave velocity $c$ plays an
important role in those mentioned universal quantities of QCR for
the 2D spin-1/2 antiferromagnets. For instance, the value of $c$
without doubt has great impact on the determinations of two universal
quantities of QCR, namely $\chi_u c^2/T \sim 0.27185$ and $c/(T\xi) \sim 1.04$.
Here $\chi_u$ and $\xi$ are the uniform susceptibility and the correlation
length, respectively \cite{Chu94,Tro98,Tan181}.
In the phase with long-range antiferromagnetic order, $c$ can be calculated
efficiently using the spatial and the temporal winding numbers
squared \cite{Jia111,Jia112,Sen15}.

Considering a clean 2D spin-1/2 AF which comes with a given spatial
arrangement of two types of antiferromagnetic couplings $J'$ and $J$
($J' > J$),
by tuning the ratio $J'/J$ (i.e. the system is dimerized) a QCP may appear when $J'/J$ exceeds a certain value
$(J'/J)_c$. The dynamic critical exponent $z$ associated with such a kind of QCP takes the value of 1.
For a QCP $g_c$ which is obtained by varying the associated parameter $g$,
the physical quantity $c$
scales as $c \propto \left(g-g_c\right)^{\nu(z-1)}$ close to $g_c$ \cite{Chu94,Tro97}, where $\nu$ is the correlation length exponent. As a
result, $c$ is a constant when the related $z$ of a QCP is 1.
For 2D quantum AF systems, the QCPs induced by dimerization introduce above
belong exactly to this case. 
When disorder is present, $z > 1$ and $c$ is zero at $g_c$. Consequently, certain universal quantities of QCR cannot be
calculated in a direct manner for disordered systems.

While for a 2D disordered quantum spin antiferromagnet,
certain quantities of QCR such as $\chi_u c^2/T$ cannot be directly accessed,
yet some observables do not encounter the difficulty that $c$ cannot be
calculated with ease. One of them, namely
the Wilson ratio $W$ \cite{Chu94,San11,Sen15}, which will be defined later,
is one of the main topics of
our study presented here. In particular, we investigate the behavior of $W$
with respect to the strength of randomness, which is controlled by a parameter $p \ge 0$,
of the employed disorder distribution. Here $p=0$ corresponds to the
clean case. Apart from $W$, the dynamic critical
exponents $z$ as well as the ground state energy densities $E_0$ of several
values of $p$ considered in this study are determined as well.

To carry out the proposed investigation, we have performed a large
scale quantum Monte Carlo calculation (QMC). In addition, several
$p$ are considered and the simulations are done at the corresponding
critical point $g_c(p)$ of each studied $p$.
Based on our numerical results, we find the magnitude of $E_0$ grows
monotonically with $g_c(p)$ (hence $p$ as well since $g_c(p) \propto p$ as shown in \cite{Pen20}), similar to that of the
correlation length exponent
$\nu$ \cite{Pen20}. Interestingly, the $p$ dependence of $z$ and $W$
are likely to be complementary to each other. For instance, while the $z$
of $0.4 \le p \le 0.9$ match well among themselves and are statistically
different from $z=1$ which corresponds to the clean system, the $W$ for
$p < 0.7$ are in reasonable good agreement with that of $p=0$ ($W \sim 0.1243$).
The subtlety of calculating these physical quantities
for a disordered system is demonstrated here as well.
Our investigation is important and interesting in itself
from a theoretical perspective. In particular, the obtained outcomes can be
used as benchmarks for future related studies.

The rest of this paper is organized as follows. After the introduction,
the studied model, the employed disorder distribution as well as the relevant
observables are described. Following
that we present our results. In particular, the numerical evidence for
the mentioned complementary relation for $z$ and $W$ are
demonstrated. Finally, a section concludes our study.

\section{Microscopic models and observables}

\begin{figure}
\vskip-0.5cm
\begin{center}
\vbox{
\includegraphics[width=0.35\textwidth]{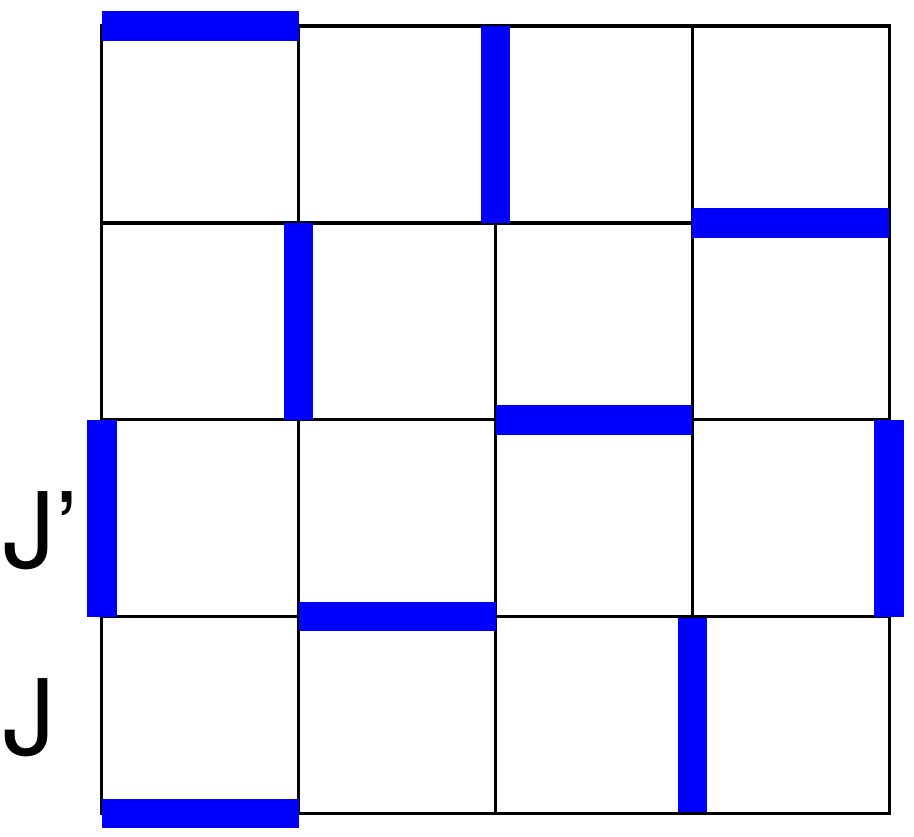}
}
\end{center}\vskip-0.5cm
\caption{The 2D dimerized spin-1/2 herringbone 
Heisenberg model on the square lattice investigated in this study. 
The antiferromagnetic 
coupling strengths for the thick and thin bonds are $J'$ and $J$, respectively.}
\label{model_fig1}
\end{figure}

The model investigated in our study has been described in detail in Refs.~\cite{Nvs14,Pen20}.
Here we briefly summarize certain technical perspectives of the considered
system. The Hamiltonian of the investigated 2D
disordered spin-1/2 herringbone Heisenberg model (on the square lattice)
is given by
\begin{eqnarray}
\label{hamilton}
H &=& \sum_{\langle ij \rangle}J\,\vec S_i \cdot \vec S_{j} 
+ \sum_{\langle i'j' \rangle}J'\,\vec S_{i'} \cdot \vec S_{j'}, 
\end{eqnarray}
where in Eq.~(1) $J$ (which are set to 1 here) and $J'$ are the
antiferromagnetic couplings (bonds) connecting nearest neighboring spins
$\langle  ij \rangle$ and $\langle  i'j' \rangle$, respectively,
and $\vec S_{i} $ is the spin-1/2 operator at site $i$. In this study
we use the convention $J' > J$.
Fig.~1 is a cartoon representation of the considered model.
The quenched disorder
introduced into the system is based on the one employed in Refs.~\cite{Nvs14,Pen20}.
Specifically, for every bold bond in fig.~\ref{model_fig1}, its
antiferromagnetic strength $J'$ takes the value of $1+(g-1)(1+p)$ 
or $1+(g-1)(1-p)$ with equal probability. Here $g>1$ and $0 \le p \le 1$.
With the used conventions, the average and difference
for these two types of bold bonds $J'$ are given by $g$ and $2p(g-1)$,
respectively. In addition, $p$ can be thought of as a measure for the disorder of the
studied model as well.

To perform the proposed calculations of determining the
ground state energy density $E_0$, the dynamic critical exponent $z$, and the Wilson ratio $W$
for the considered disordered system (with various $p$),
the uniform susceptibility $\chi_u$, the internal energy density $E$, 
and the specific heat $C_V$ (as functions of the temperature $T$ or the inverse temperature $\beta$)
are measured in our simulations.

The uniform susceptibility $\chi_u$ is defined by
\begin{equation}
\chi_u = \frac{\beta}{L^2}\Bigg\langle \left( \sum_{i}S_i^z\right)^2 \Bigg\rangle, 
\end{equation} 
where $\beta$ and $L$ are the inverse temperature and linear box size
used in the simulations, respectively.
Furthermore, the internal energy density $E$ and the specific heat $C_V$ are given as 
\begin{eqnarray}
E = \frac{1}{L^2}\langle H \rangle, \\ 
C_V = \frac{\partial E }{\partial T}. 
\end{eqnarray}

Using these observables, $E_0$, $z$, and $W$ for various $p$ of the studied
disordered model can be determined with high precision.

\section{The numerical results}

For each of the considered values of $p=$ 0.0, 0.2, 0.4, 0.5, 0.6, 0.7, 0.8, and 0.9, to calculate the
associated desired physical quantities, 
we have carried out large-scale QMC
using the stochastic series expansion (SSE) algorithm with very efficient 
operator-loop update \cite{San99,San10}. The simulations
are done at the corresponding critical points $g_c(p)$ for these chosen $p$.
In addition, for every
$p$, several hundred to
few thousand randomness configurations with $L=128$ and (or) $L=256$ are generated.
Each configuration is produced with its own random seed and then is used
for all the calculations of the considered values of $\beta$.

\subsection{The strategy of calculating the Wilson ratio $W$}

In the framework of SSE,
the quantities of internal energy density $E$ and and specific heat $C_V$ can
be obtained by
\begin{eqnarray}
  E &=& -\frac{1}{L^2}\left(\langle n \rangle/\beta - \frac{1}{4}\sum_{b}J_b \right), \\ 
  C_V &=& \frac{1}{L^2}\left(\langle n^2\rangle - \langle n \rangle^2 - \langle n \rangle\right), 
  \end{eqnarray}
respectively, where the summation is over all the bonds $b$ and $n$ is the number of nonidentity operators in the
SSE operators sequence (operators string).

Based on the large-$N$ expansion of the relevant effective field theory,
at the associated critical point it is predicted that for clean systems the (leading) low-$T$ behavior of
$\chi_u$, $E$, and $C_V$ are given by
\begin{eqnarray}
&&\chi_u \sim \frac{1.0760}{\pi c^2}T,\\
&&E \sim E_0 + \frac{2.8849}{\pi c^2}T^3, \\
&&C_V \sim \frac{8.6548}{\pi c^2}T^2,
\end{eqnarray}  
respectively, where the $c$ appearing above is the spinwave velocity. With these
leading $T$-dependence of $\chi_u$, $E$ and $C_V$, the Wilson ratio $W$ can be
expressed as 
\begin{eqnarray}
W = \frac{\chi_uT}{C_V} \sim 0.1243.
\label{wr}
\end{eqnarray}

While $C_V$ can be calculated directly from its definition 
$C_V = \partial E/\partial T$ (or $C_V = \frac{1}{L^2}\left(\langle n^2\rangle - \langle n \rangle^2 - \langle n \rangle\right)$),
as being shown in the literature, such a approach will lead to very noisy results at the
region of low temperature \cite{Sen15}. 
In addition, the fact that $c$ is zero at the QCP of a disordered system
prevents one from determining $c$ directly.
Motivated by the method outlined in Ref.~\cite{Sen15}, 
here we calculate $W$ through the following procedures.

Firstly, from the $\beta$-dependence of the internal energy density $E$, namely

\begin{eqnarray}
  E(\beta) = E_0 + a\beta^{-1-2/z}
\label{energy}\end{eqnarray}
(here $z$ is
the dynamic critical exponent),
one obtains $a$ and $z$. Then the specific heat $C_V$, as a function of $\beta$,
can be written as

\begin{eqnarray}
C_V = a(1+2/z)\beta^{-2/z}.
\label{specific_heat}
\end{eqnarray}

Secondly, the $\beta$-dependence of $\chi_u$ is fitted
to the expression 
\begin{eqnarray}
\chi_u(\beta) = b\beta^{1-2/z}.
\label{chiu}
\end{eqnarray} 

Finally, using Eqs.~\ref{specific_heat} and \ref{chiu},
one arrives at the following formula for $W$
\begin{eqnarray}
  W = \frac{b}{a(1+2/z)}.
  \label{wilson_ratio}
\end{eqnarray}

In other words, instead of using $C_V$ directly, here $W$ is calculated through the
coefficients $z$, $a$ and $b$ obtained
from fitting the data of $\chi_u$ and $E$ to their expected $T$-dependence ansatzes Eqs.~\ref{energy} and \ref{chiu}.

\subsection{The obtained $\chi_u$ and $E$ from simulations}

\begin{figure}
\vskip-0.5cm
\begin{center}
  \vbox{
    \includegraphics[width=0.4\textwidth]{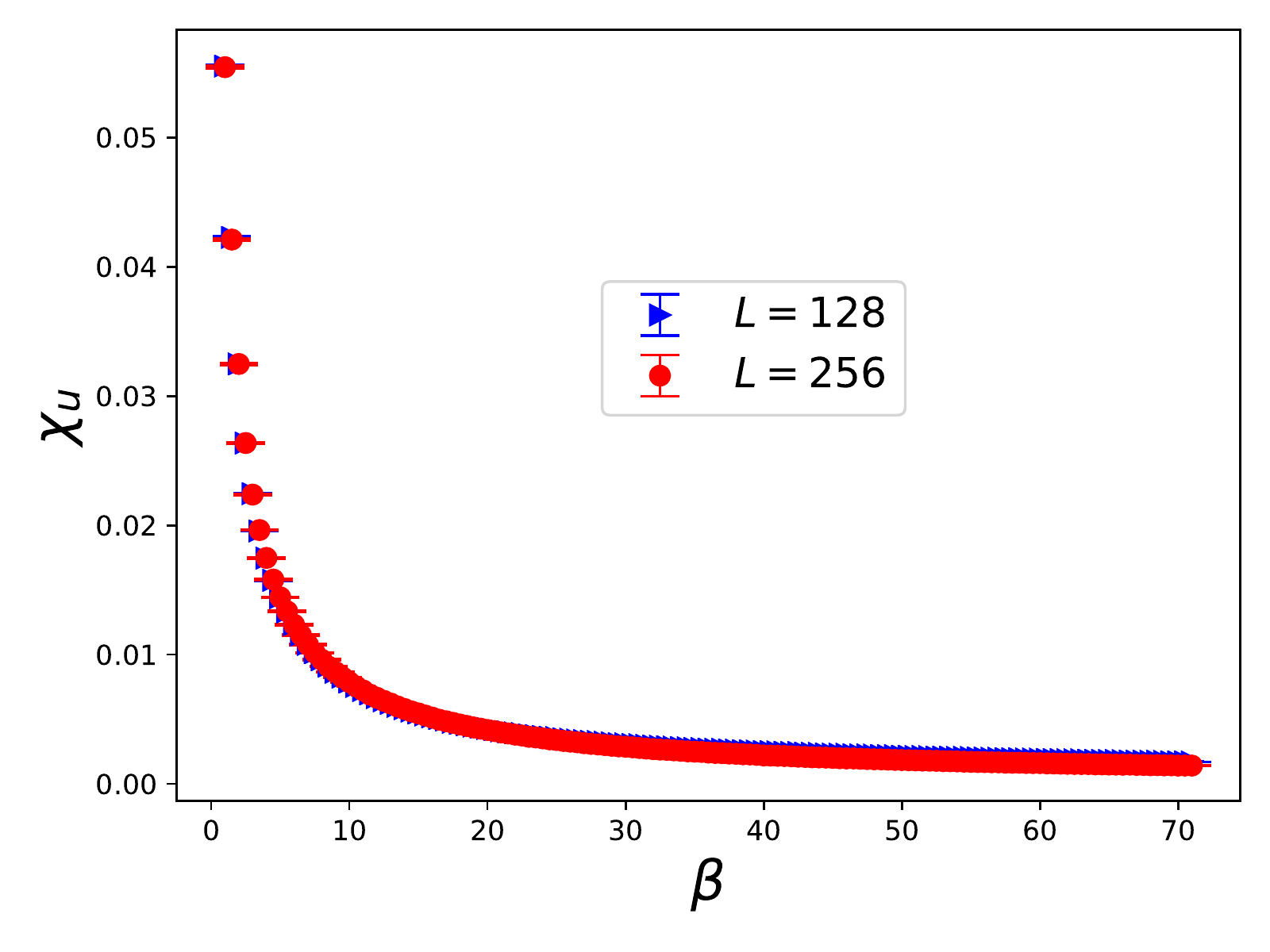}\vskip0.25cm
\includegraphics[width=0.4\textwidth]{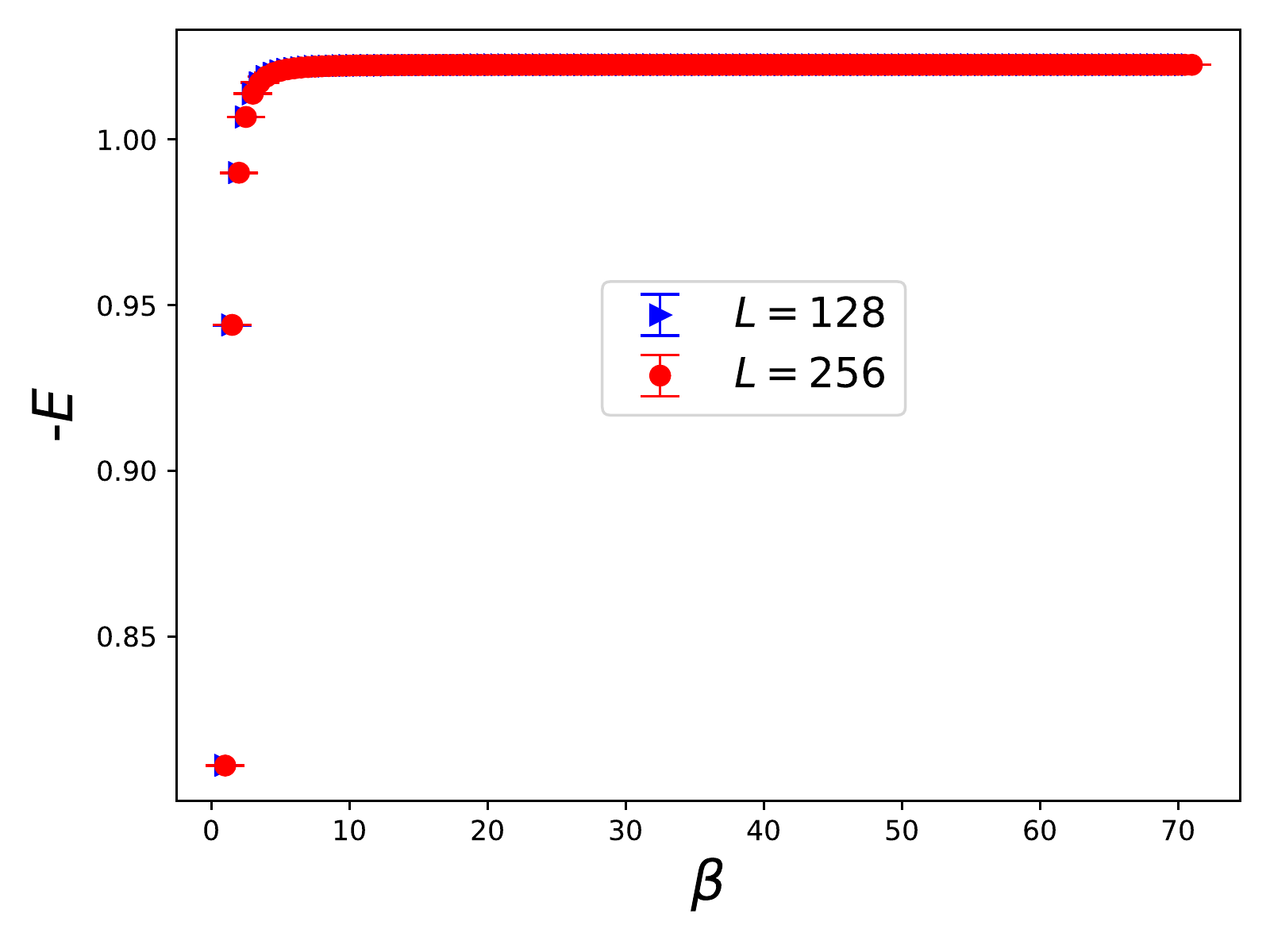}
}
\end{center}\vskip-0.5cm
\caption{$\chi_u$ (top) and minus (internal) energy density $E$ (bottom) as functions
  of $\beta$ for $p=0.0$. The shown errors are the corresponding mean errors.}
\label{results_p00}
\end{figure}

\begin{figure}
\vskip-0.5cm
\begin{center}
\vbox{\includegraphics[width=0.4\textwidth]{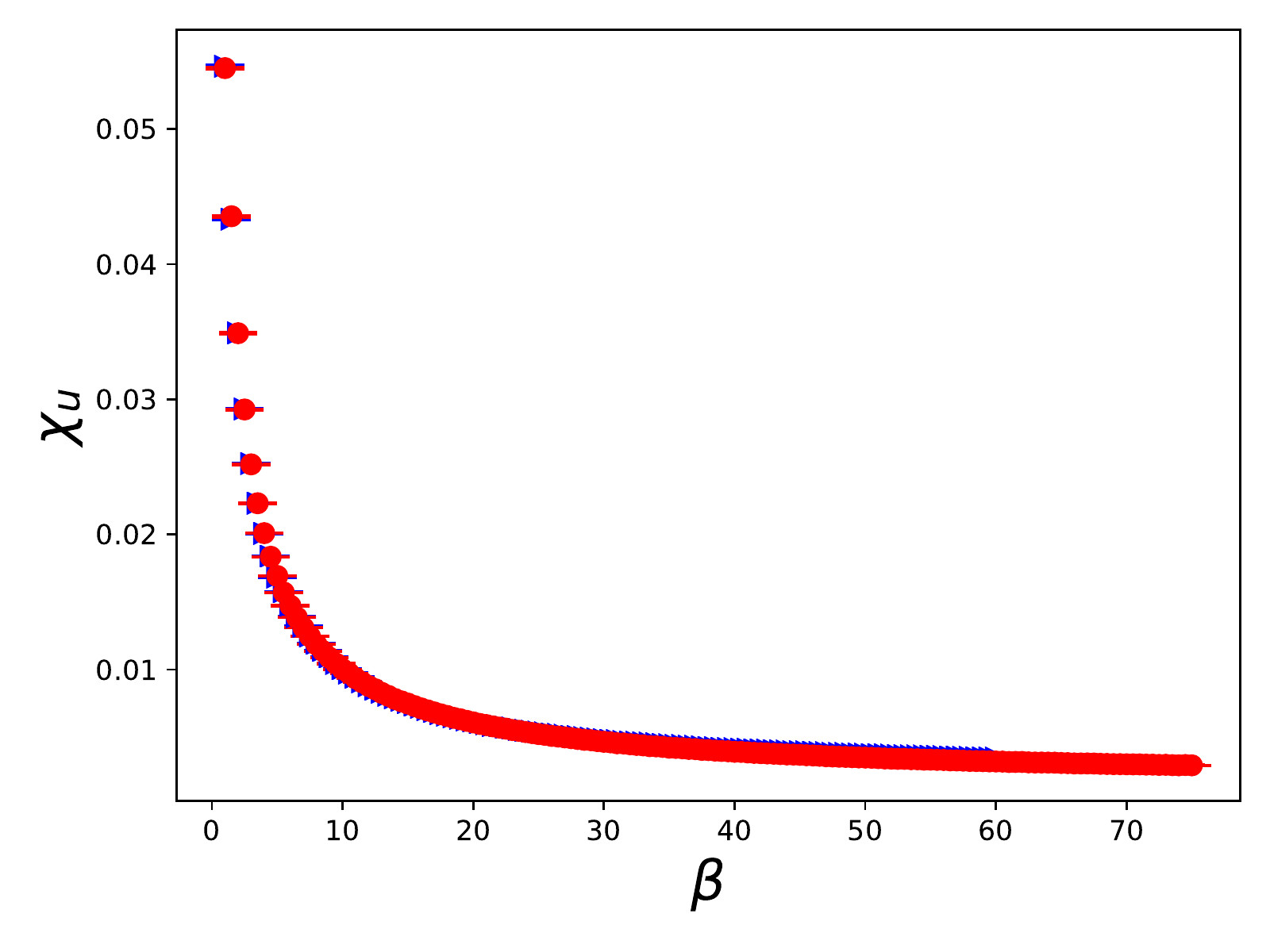}\vskip0.25cm
\includegraphics[width=0.4\textwidth]{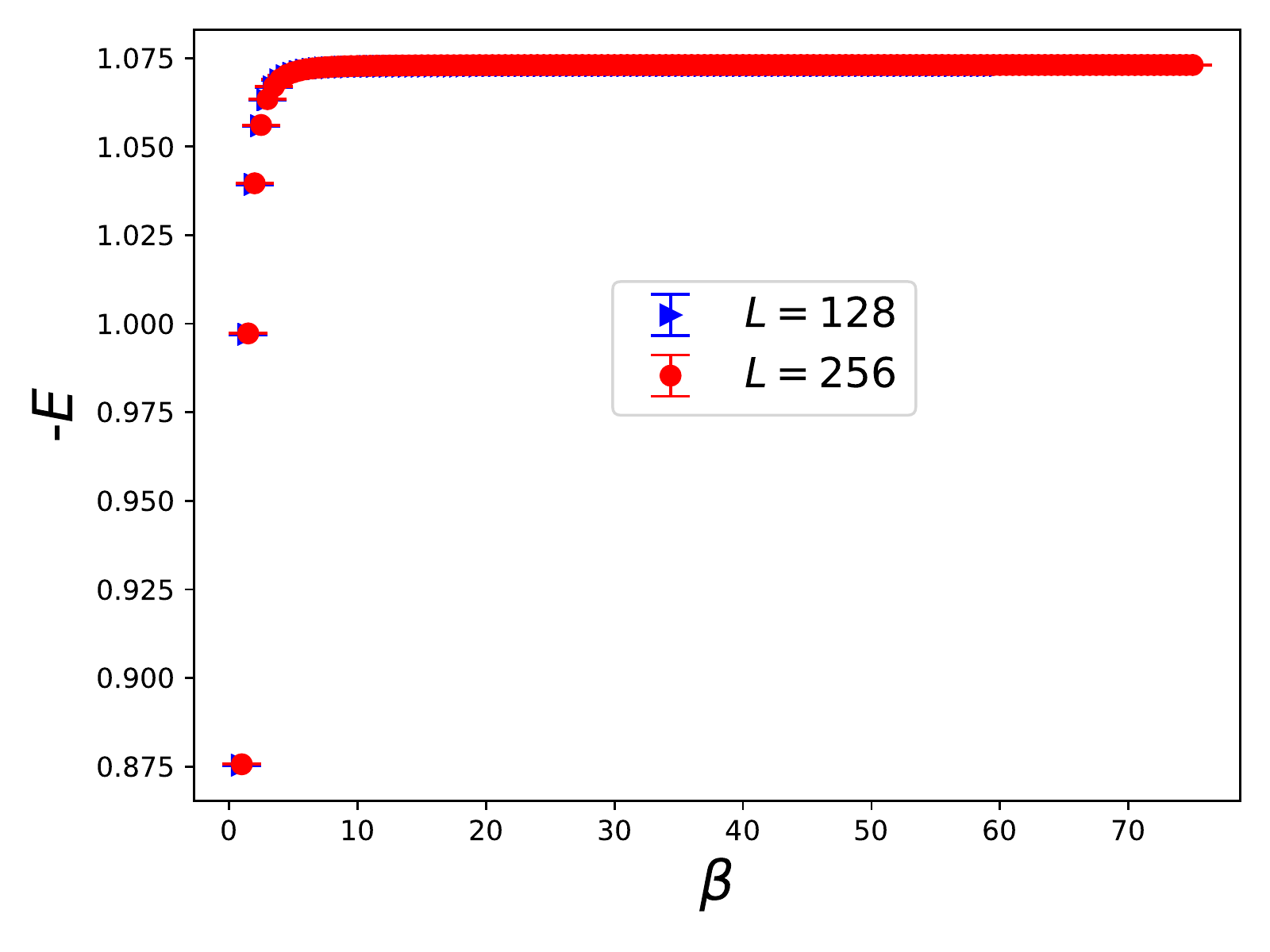}}
\end{center}\vskip-0.5cm
\caption{$\chi_u$ (top) and minus (internal) energy density $E$ (bottom) as functions of
  $\beta$ for $p=0.4$. The shown errors are the corresponding mean errors.}
\label{results_p04}
\end{figure}

\begin{figure}
\vskip-0.5cm
\begin{center}
  \vbox{
    \includegraphics[width=0.4\textwidth]{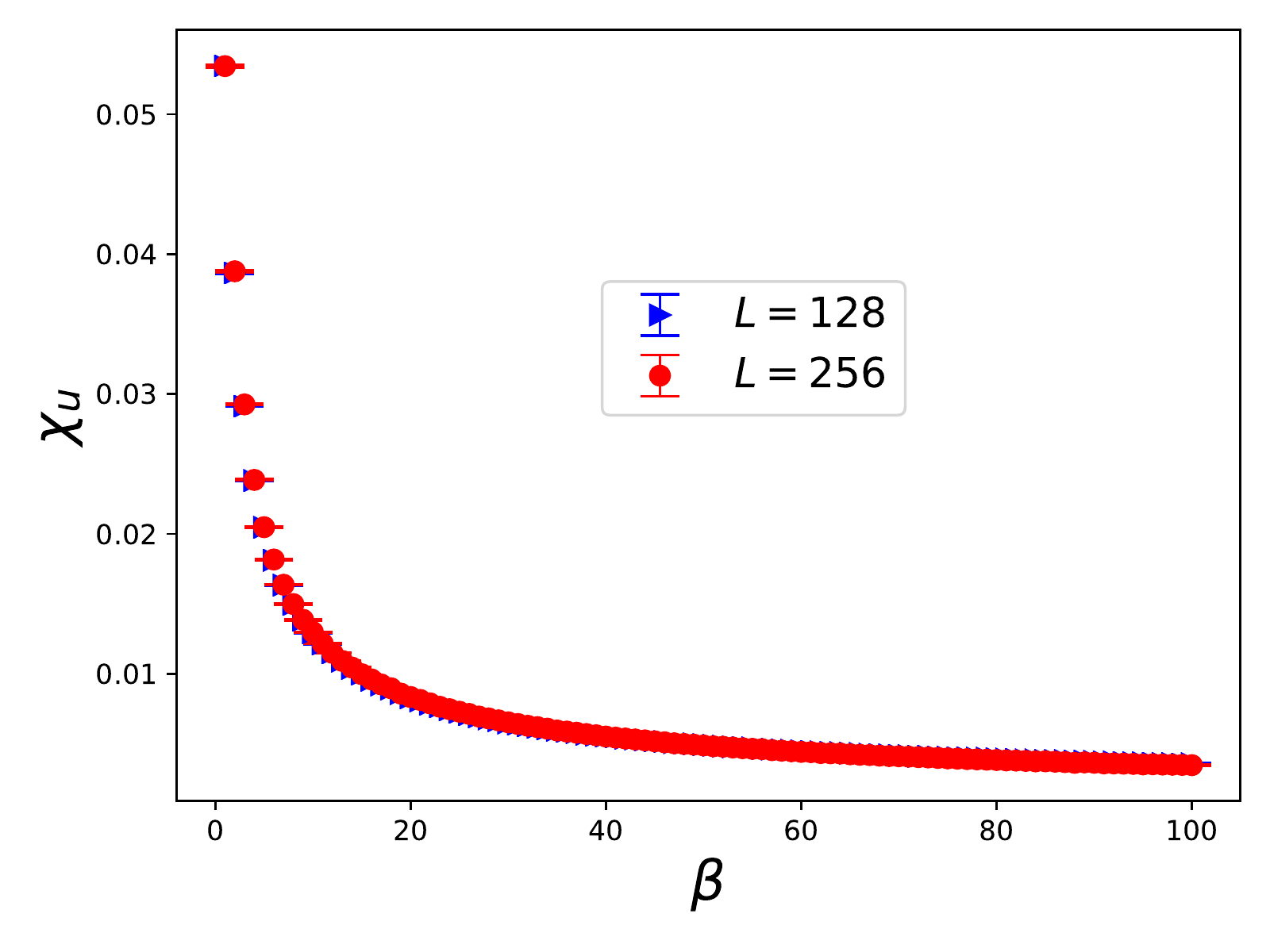}\vskip0.25cm
\includegraphics[width=0.4\textwidth]{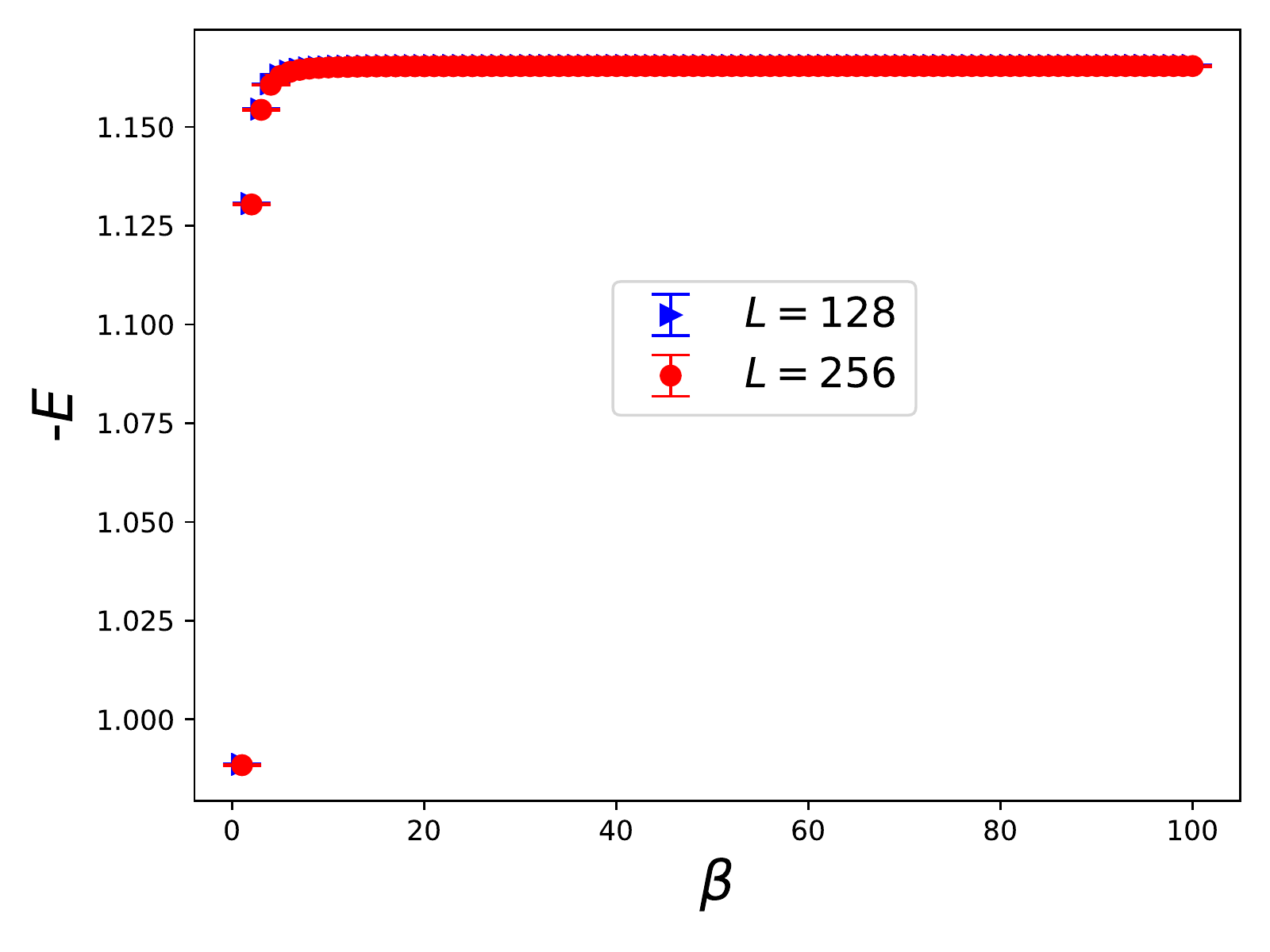}
}
\end{center}\vskip-0.5cm
\caption{$\chi_u$ (top) and minus (internal) energy density $E$ (bottom) as functions of
  $\beta$ for $p=0.6$. The shown errors are the corresponding mean errors.}
\label{results_p06}
\end{figure}

\begin{figure}
\vskip-0.5cm
\begin{center}
  \vbox{
    \includegraphics[width=0.4\textwidth]{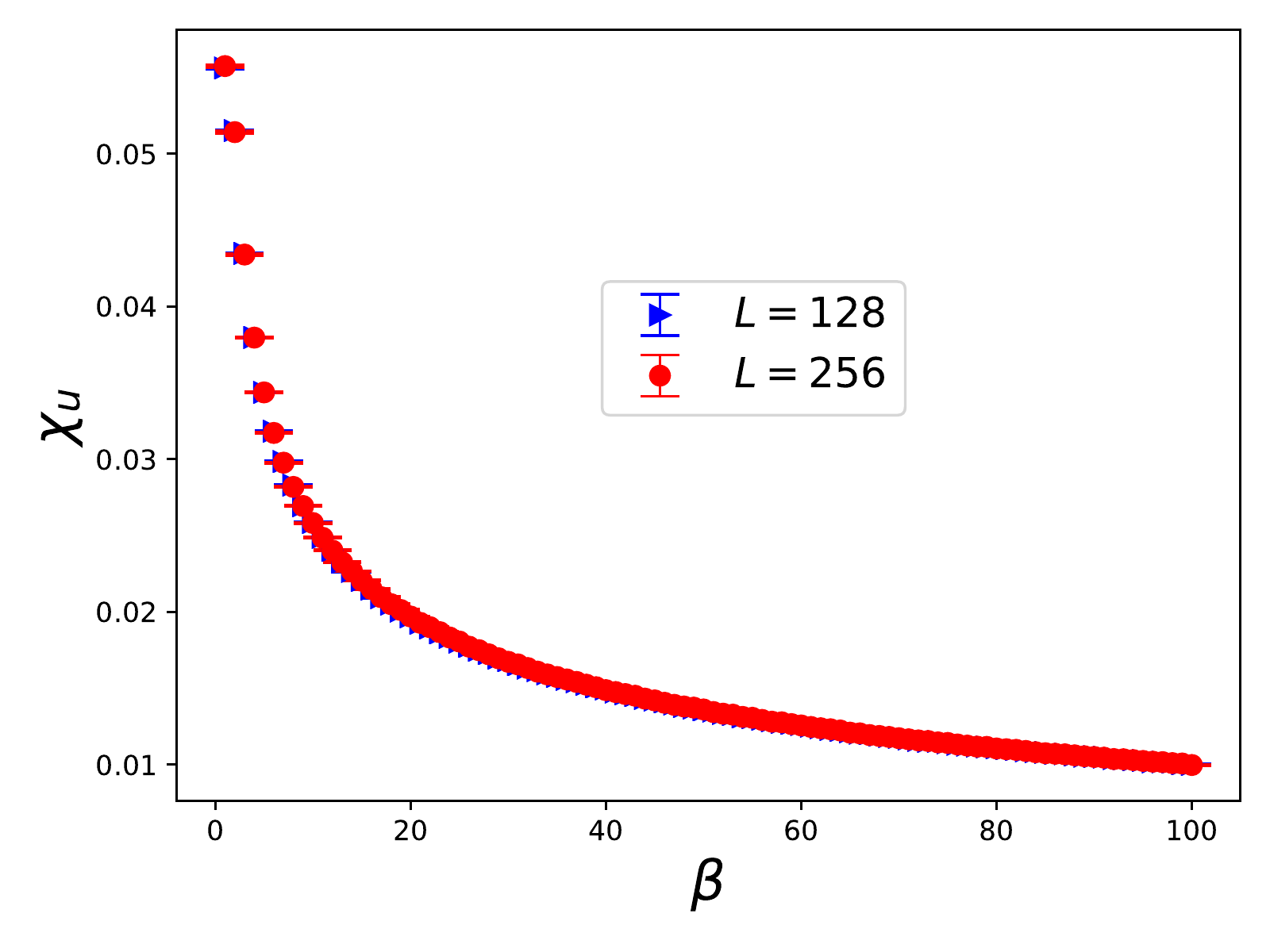}\vskip0.25cm
\includegraphics[width=0.4\textwidth]{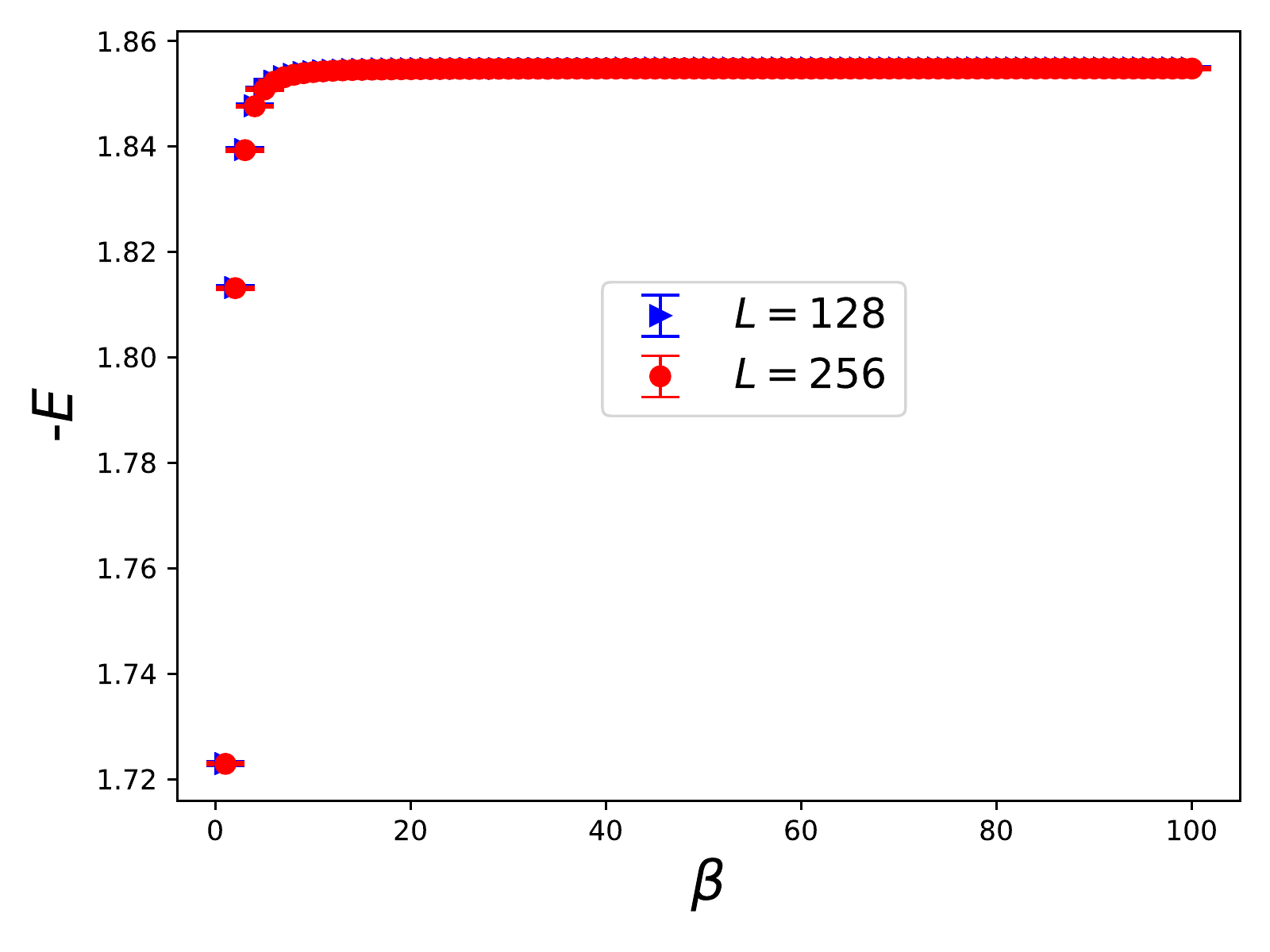}
}
\end{center}\vskip-0.5cm
\caption{$\chi_u$ (top) and minus (internal) energy density $E$ (bottom) as functions of
  $\beta$ for $p=0.9$. The shown errors are the corresponding mean errors.}
\label{results_p09}
\end{figure}

The obtained data of $\chi_u$ and $-E$ for $p=0.0$, 0.4, 0.6, and 0.9 are depicted in
figs.~\ref{results_p00},~\ref{results_p04},~\ref{results_p06}
and \ref{results_p09}.
Both data of $L=128$ and $L=256$
are put in these figures in order to demonstrate that the outcomes of $L=256$
are (most likely) sufficient
for size convergence.

\subsection{The analysis results associated with the clean model}

For the clean model $p=0$, it is well known that $z = 1$. Hence, $z$ will
be fixed to 1 in our analysis for $p=0$. As a result,
the following ansatz 
\begin{equation}
\chi_u = a_0 + aT + a_1T^{2}
\end{equation}
is considered to fit the $\chi_u$ data of $p=0$. Apart from that, the formula
used to fit the data of $E$ for $p=0$ is Eq.~\ref{energy} with $z = 1$ as well.

By investigating the relevant data for $L=128$ and $L=256$, the finite size effect begins to
appear when $\beta > 16.0$. Therefore the data of $L=256$ with $\beta \le 24.0$ are used for
the fits. We have additionally carried out fits using the $L=256$ data of $\beta \le 20.0$
and have found that these new results lead to a value of $W$ which agrees quantitatively
with that obtained using the $L=256$ data of $\beta \le 24.0$.
For each of data with a fixed range of $\beta$, in additional to taking care of finite
size effect, the following procedures are adopted to calculate the corresponding $W$.

Firstly, a bootstrap resampling (with respect to $\beta$) is conduct simultaneously
for both $\chi_u$ and $E$. Secondly, fits
for these obtained resampled data are performed. Finally $W$ is determined using the
outcomes of these fits.
In these mentioned fits, Gaussian noises are considered as well. The above described steps are carried out for twenty thousand
times and only those outcomes with both $\chi^2$/DOF (DOF stands for degrees of freedom)
of the two fits (for $\chi_u$ and $E$) being smaller than 3.0 are included as the candidate
results of $W$.
The resulting $W$, as well as its associated uncertainty, quoted for this set of data with that given fixed
range of $\beta$ are the mean and the standard deviation of these candidate results.

We have conducted several calculations using various range of $\beta$, and each of these
calculations comes with is own results
(mean and uncertainty) for $W$. Moreover, to estimate the means and errors of the desired
quantities appropriately,
the weighted bootstrap resampling method is
applied to all the mentioned results of $W$.
Specifically, for every randomly generated data set $(W_j,\sigma_{W_j})$ obtained
using the bootstrap procedure ($\sigma_{W_j}$ is the standard deviation
associated with $W_{j}$), the resulting mean is given by

\begin{eqnarray}
\frac{\sum_i \frac{1}{\sigma_{W_{i}}^2}W_{i}}{\sum_i\frac{1}{\sigma_{W_{i}}^2}}.
\end{eqnarray}

The reason for the use of above equation (called weighted mean in this study)
is as follows. Notice that data with large standard deviations are less
accurately determined than those with small standard deviations.
As a result, those large standard deviation data should contribute
less weight to the determination of the associated mean.

After carrying out twenty thousand weighted bootstrap resampling steps, the resulting $W$
is estimated to be $W = 0.1238(3)$. The obtained $W=0.1238(3)$ agrees very well with the
theoretical prediction $W=0.1243$.
This confirms the validity of the procedures introduced above for the calculation of $W$.

The ground state energy density for the clean model is calculated by the same procedure
and is given by $E_0 = -1.022523(1)$.

\subsection{The results of the disordered model with various randomness strength $p$}

Since each generated configuration is used for all the simulations of the
considered $\beta$, for a given set of $p > 0$ and $L$, the data themselves are correlated. Hence,
to accurately estimate the associated errors for the coefficients in the fitting ansatzes,
one should employ the correlated least $\chi^2$ method for the analysis.
However, the stability of the correlated least $\chi^2$ method
varies and depends on the quality of the data used for the fits. Moreover, biased outcomes may be reached
if the associated covariance matrix for a given data set contains eigenvalues which have
very small magnitude. Using the rule of thumb that the ansatz considered to fit correlated
data should contain as few (to be determined) parameters as possible, we adopt
the following approach to calculate $E_0$, $z$, and $W$ for $p > 0$.

First of all, for each $\beta$ the bootstrap resampling method is performed for the raw data resulting from the generated disordered
configurations.
Second of all, these resampled data are used to calculate the disordered average of $\chi_u$ and $E$ which are then
considered for the relevant fits. Here the data employed for the fits of $\chi_u$ and $E$ have different range of $\beta$.
Indeed, as can be seen from figs.~\ref{results_p04}, \ref{results_p06}, \ref{results_p09}, for all the considered $p > 0$
the associated $E$ reaches its ground state value $E_0$ quickly, while this is not the case for $\chi_u$. As a result, it is
more appropriate to use different range of $\beta$ for the fits of $\chi_u$ and $E$.

After carrying out the fit of $\chi_u$, the obtained
result of $z$ is employed as an input for the fit of $E$. When both fits of $\chi_u$ and
$E$ are done, the resulting results are then
put back to calculate the associated correlated $\chi^2$/DOF. Here a cut-off for the
eigenvalues of the associated covariance matrix is imposed in order to avoid biased results.
These introduced steps are
performed for many times, and
only those results which have correlated $\chi^2$/DOF smaller than 3.0
for both the fits of $\chi_u$ and $E$ are considered for later calculations.
Finally for each of the considered $p$, the above procedures
have been applied to many sets consisting of various range of $\beta$. Each of these
sets (The whole sets is denoted by $S$) has its own results (mean and standard deviation) of $E_0$, $z$,
and $W$ as well as the number of successful calculations. 

For a considered $p$, the final quoted results of $E_0$, $z$, and $W$ in this study are
estimated by a bootstrap resampling procedure using the
following formula to calculate the mean of every resampled data from $S$.

\begin{eqnarray}
\frac{\sum_i N_i O_i /\sigma_{O_i}^2}{\sum_i N_i/\sigma_{O_i}^2},
\end{eqnarray}
where $\{O_i\}$, $\{\sigma_{O_i}\}$ and $\{N_i\}$ stand for the randomly picked outcomes in $S$, the associated
standard deviations, as well as the related numbers of
(successful calculated) results of these picked outcomes, respectively.
Finally, such a resampling step is conducted for several thousand times, and the numerical values
presented here for these considered
physical quantities are the resulting means and standard deviations (estimated conservatively)
of this procedure. The uncertainties of $E_0$ calculated by the described
steps are much smaller than those of the original $E_0$ contained in $S$.
Hence, for the data in $S$ we have also calculated their associated weighted errors. 
The dominant one of these two estimations, namely the standard deviations and the weighted errors, 
are the final values quoted here.

\begin{figure}
\vskip-0.5cm
\begin{center}
\includegraphics[width=0.4\textwidth]{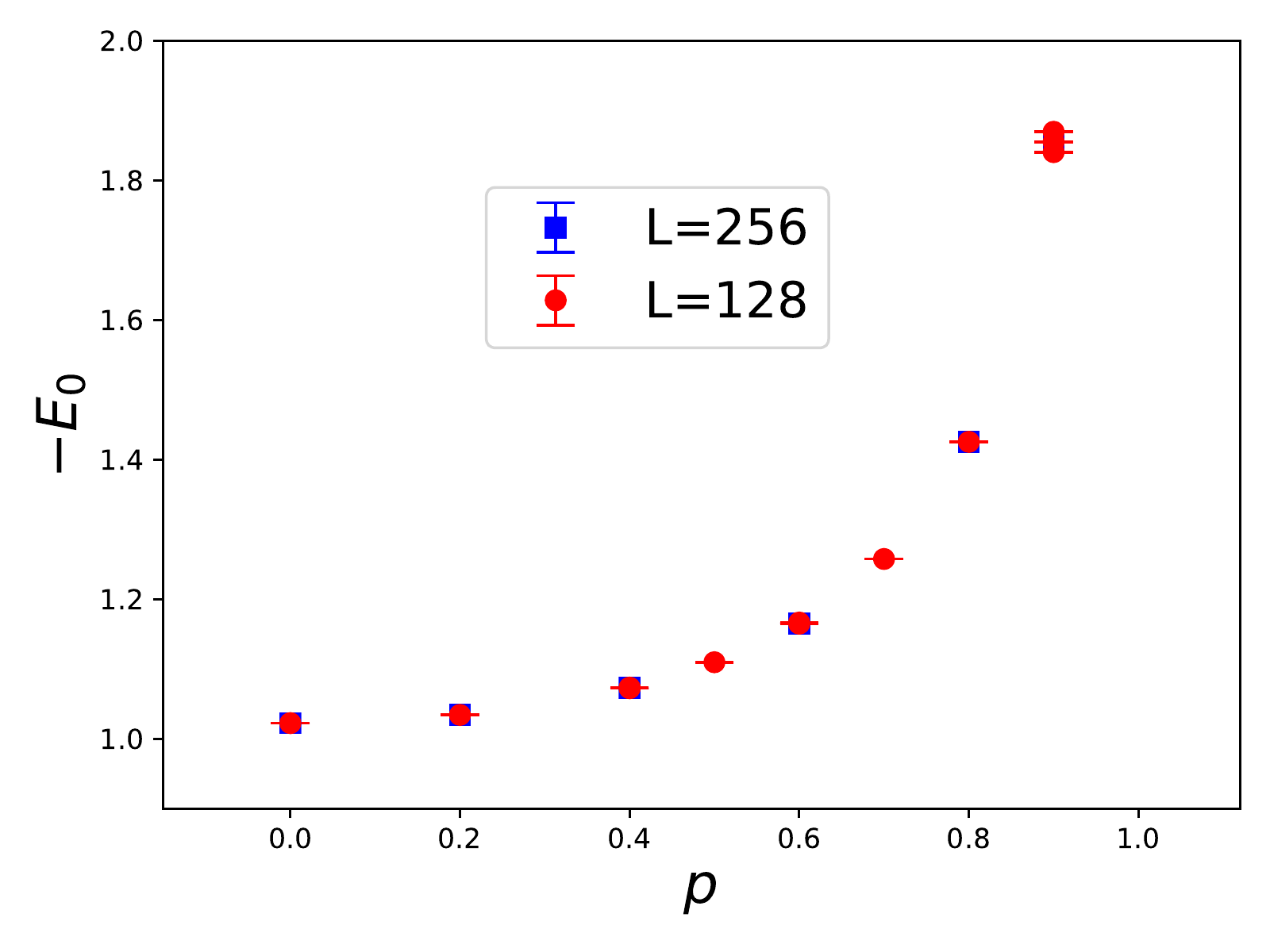}
\end{center}\vskip-0.5cm
\caption{-$E_0$ as functions of $p$. The results are obtained from the analysis using the correlated $\chi^2$
  described in the main text. The solid squares and solid circles are for $L=256$
  and $L=128$, respectively. For some values of $p$, the $L=128$ results contains
  those corresponding to $g_c$, the lower and upper bounds of $g_c$. 
 }
\label{figE_0}
\end{figure}

\begin{figure}
\vskip-0.5cm
\begin{center}
\includegraphics[width=0.4\textwidth]{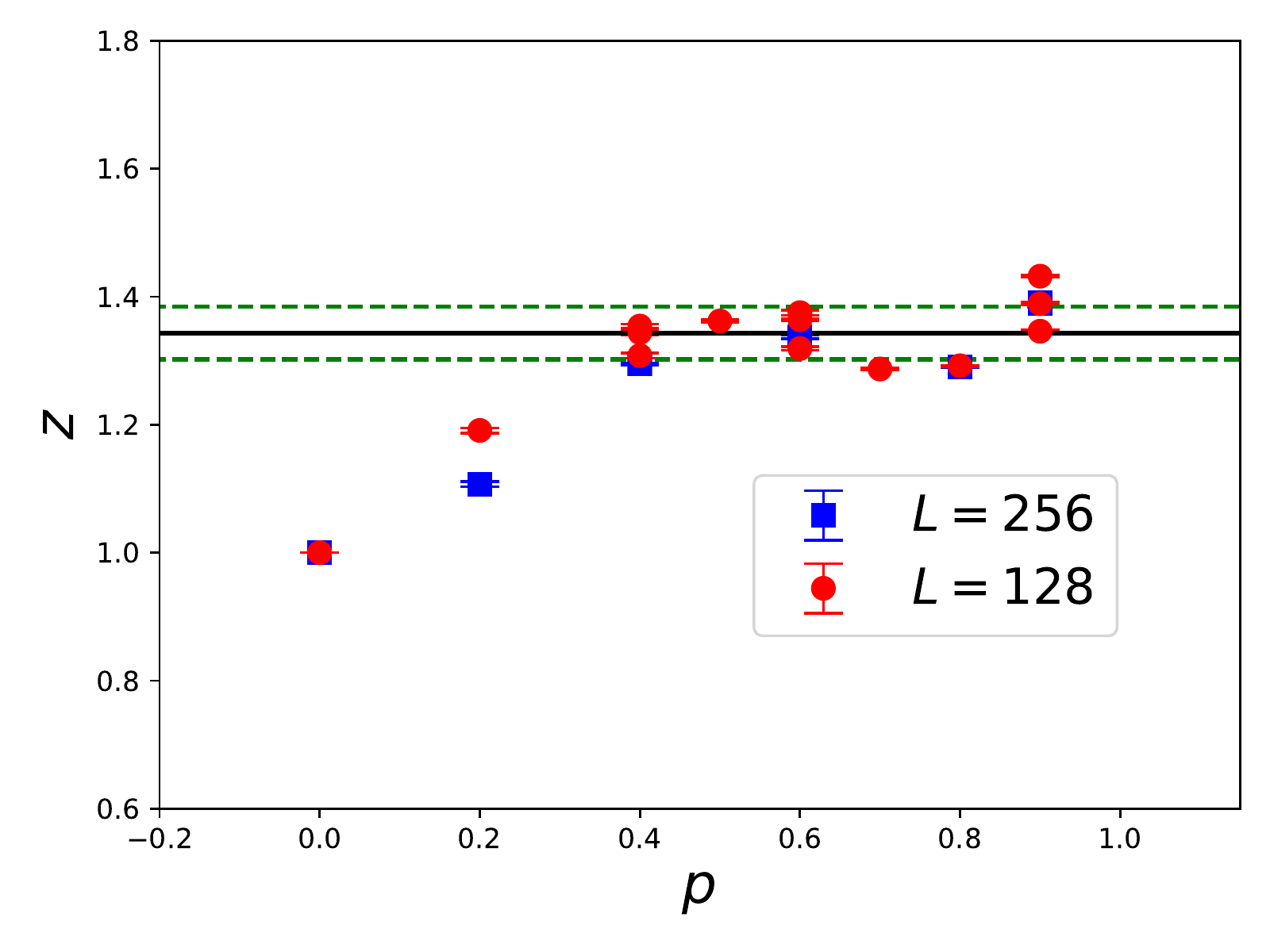}
\end{center}\vskip-0.5cm
\caption{$z$ as functions of $p$. The results are obtained from the analysis using the correlated $\chi^2$
  described in the main text. The solid squares and solid circles are for $L=256$
  and $L=128$, respectively. For some values of $p$, the $L=128$ results contains
  those corresponding to $g_c$, the lower and upper bounds of $g_c$. The solid and dashed lines in the figure
  represent the mean and standard deviation for all the $z$ such that their
  associated $p$ satisfy $p \ge 0.4$ (including both those of $L=128$ and $L=256$).
 }
\label{figz}
\end{figure}

\begin{figure}
\vskip-0.5cm
\begin{center}
\includegraphics[width=0.4\textwidth]{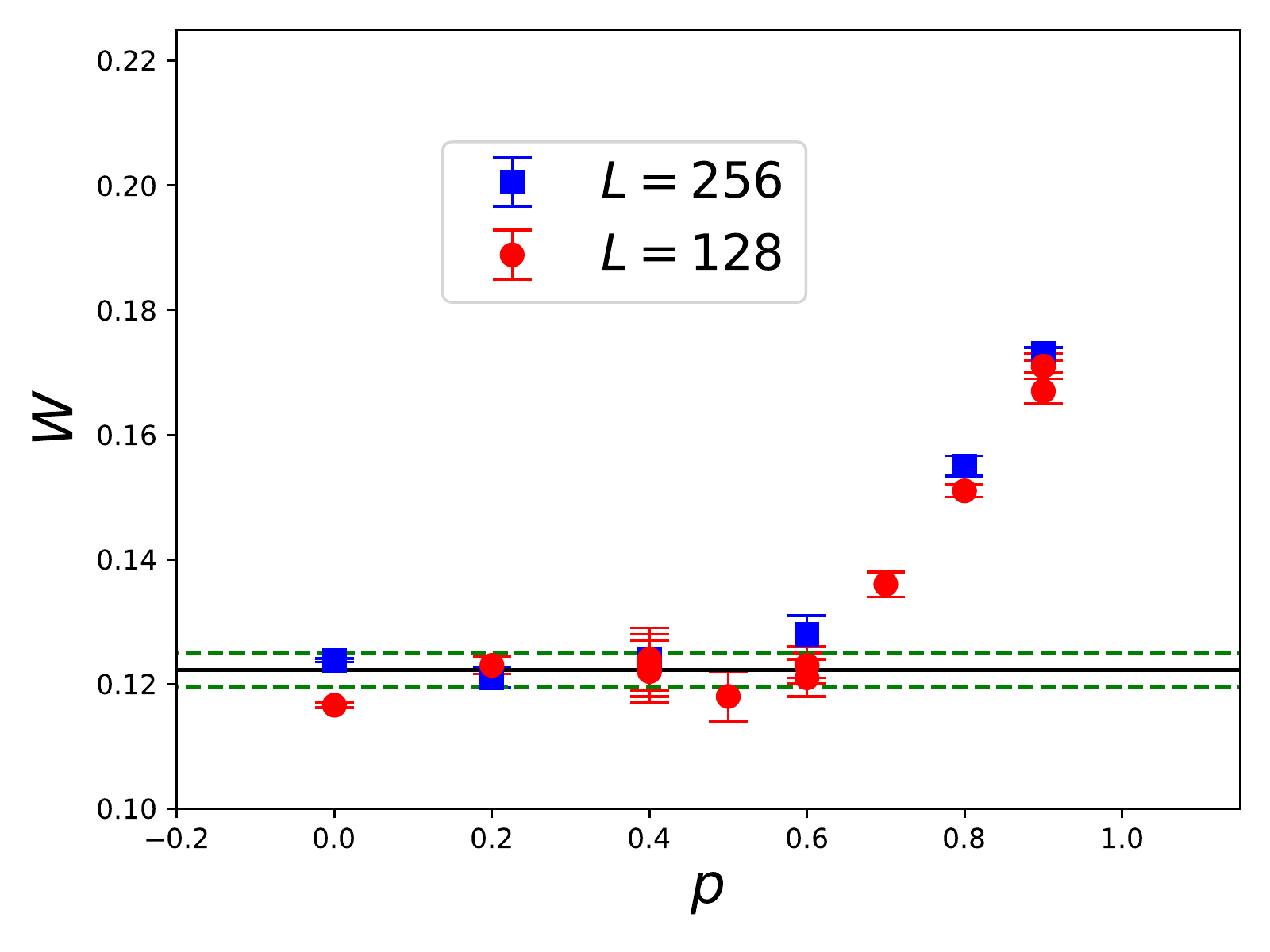}
\end{center}\vskip-0.5cm
\caption{$W$ as functions of $p$. The results are obtained from the analysis using the correlated $\chi^2$
  described in the main text.
  The solid squares and solid circles are for $L=256$
  and $L=128$, respectively. For some values of $p$, the $L=128$ results contains
  those corresponding to $g_c$, the lower and upper bounds of $g_c$. The solid and dashed lines in the figure
  represent the mean and standard deviation for all the $W$ such that
  their corresponding $p$ satisfy $p < 0.7$ (including both those of $L=128$ and $L=256$). 
 }
\label{figW}
\end{figure}

The $E_0$, $z$ and $W$ as functions of $p$ calculated by the procedures introduced above
are shown in figs.~\ref{figE_0},~\ref{figz}~,\ref{figW}.
The related results for the clean model are shown in these figures as well for comparison.

The $E_0$ as a function of $p$ shows a monotonic behavior in magnitude, which is similar to
that of the correlation length exponents $\nu$ obtained in Ref.~\cite{Pen20}.

Regarding the $z$ presented in the figure, one observes that the magnitude
of $z$ increases with $p$ until $p$ reaches a specific $p_c < 0.4$.
For $p \ge 0.4$, all
the calculated values of $z$ lie between (around) 1.3 and (around) 1.4.
If one takes into
account the systematic errors due to the uncertainties of $g_c(p)$, then the $z$ for $p \ge 0.4$
are fairly close to each other. The solid and dashed lines in fig.~\ref{figz} represent the
mean and standard deviation for all the values of $z$ associated with $p \ge 0.4$ (including both those of $L=128$ and $L=256$). These guided lines justify the claim
made above.
This phenomenon is consistent with the one shown in Ref.~\cite{Skn04},
where the calculated $z$ corresponding to various parameters take a universal value.
We would like to point out that when conducting the determination of $z$ from $\chi_u$,
the obtained results are somehow a little bit sensitive to the considered fitting range of $\chi_u$.
This motives the use of the resampling procedures described above. In conclusion, our analysis indicates that it
is subtle to calculate the quantity $z$ and a careful strategy is needed.

Finally, in fig.~\ref{figW} we demonstrate the results of $W$ as functions of $p$ obtained from
the analysis outlined previously. Intriguingly, similar to the scenario of $z$, for those $W$ corresponding to $p < 0.7$,
their values are more or less close to each other. The solid and dashed lines in the figure
again stand for the mean and standard deviation for all the $W$ with their associated $p$ satisfy $p < 0.7$ (including both those of $L=128$ and $L=256$).
Considering the impact resulting from the errors of $g_c(p)$, the scenario that $W$ take the same
value (or at least values close to each other) for all the $p$ such that
$p < 0.7$ is probable. 
Interestingly, the correlation length exponent $\nu$ is beginning to fulfill the Harris criterion
when $p > 0.8$ and this is where the magnitude of $W$ increases sharply. This observation
implies that there may exist a relation between $W$ and fulfillment of Harris criterion.

\section{Discussions and Conclusions}

\begin{figure}
\vskip-0.5cm
\begin{center}
\includegraphics[width=0.4\textwidth]{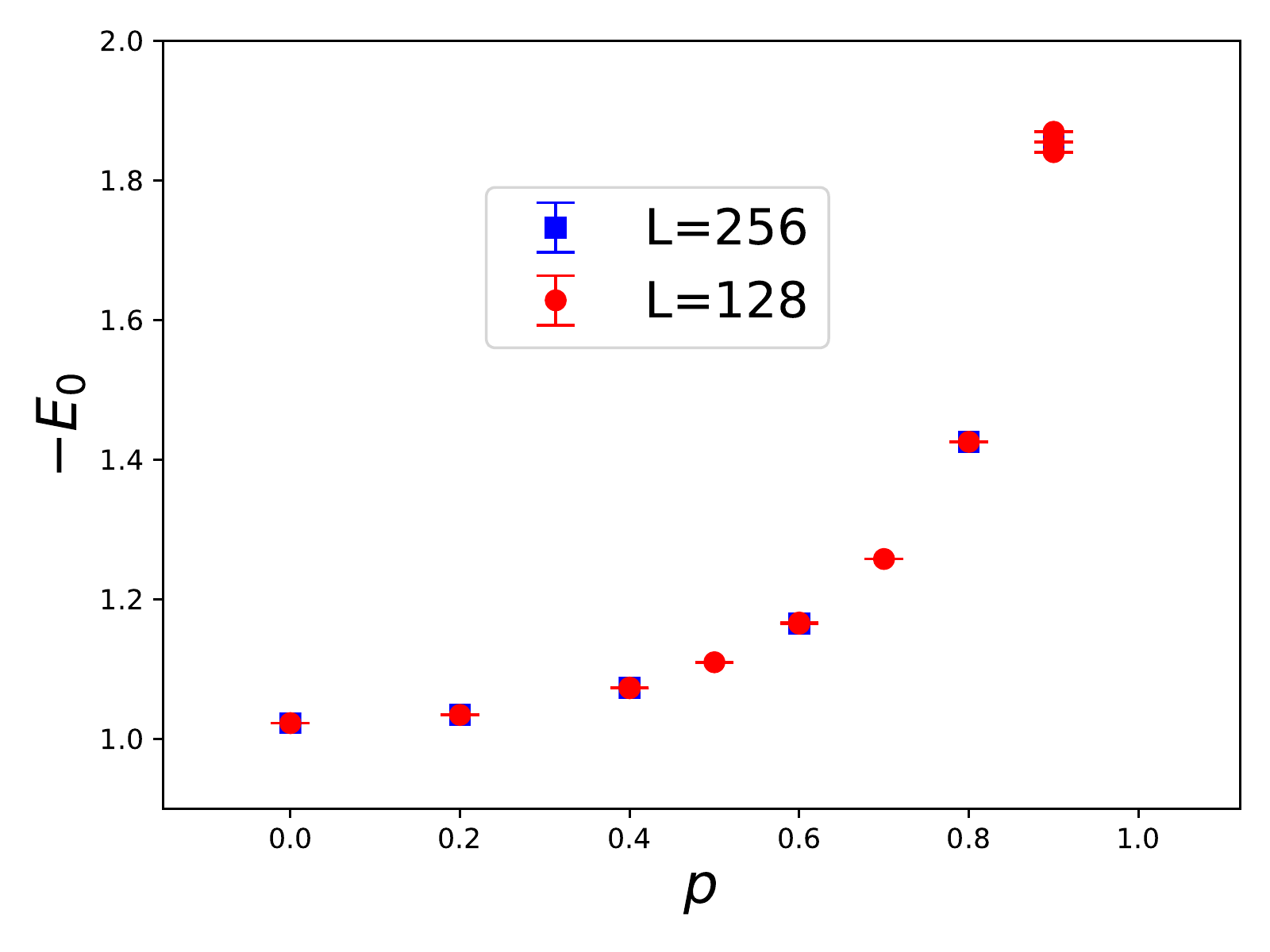}
\end{center}\vskip-0.5cm
\caption{-$E_0$ as functions of $p$. The results are obtained from the analysis using the conventional uncorrelated $\chi^2$.
  The solid squares and solid circles are for $L=256$
  and $L=128$, respectively. For some values of $p$, the $L=128$ results contains
  those corresponding to $g_c$, the lower and upper bounds of $g_c$. 
 }
\label{cfigE_0}
\end{figure}

\begin{figure}
\vskip-0.5cm
\begin{center}
\includegraphics[width=0.4\textwidth]{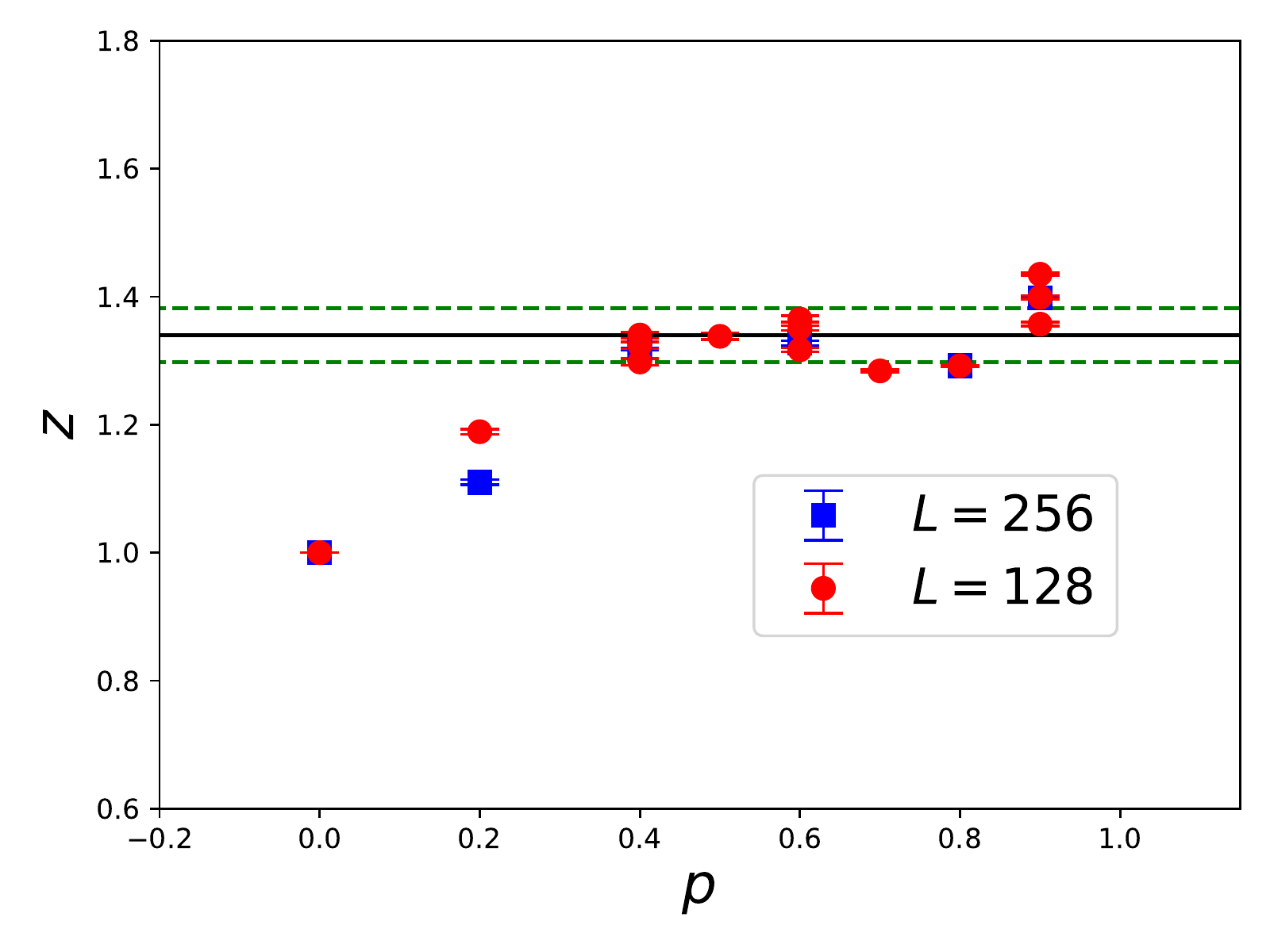}
\end{center}\vskip-0.5cm
\caption{$z$ as functions of $p$. The results are obtained from the analysis using the conventional uncorrelated $\chi^2$.
  The solid squares and solid circles are for $L=256$
  and $L=128$, respectively. For some values of $p$, the $L=128$ results contains
  those corresponding to $g_c$, the lower and upper bounds of $g_c$. The solid and dashed lines in the figure
  represent the mean and standard deviation for all the $z$ such that their
  associated $p$ satisfy $p \ge 0.4$ (including both those of $L=128$ and $L=256$).
 }
\label{cfigz}
\end{figure}

\begin{figure}
\vskip0.5cm
\begin{center}
\includegraphics[width=0.4\textwidth]{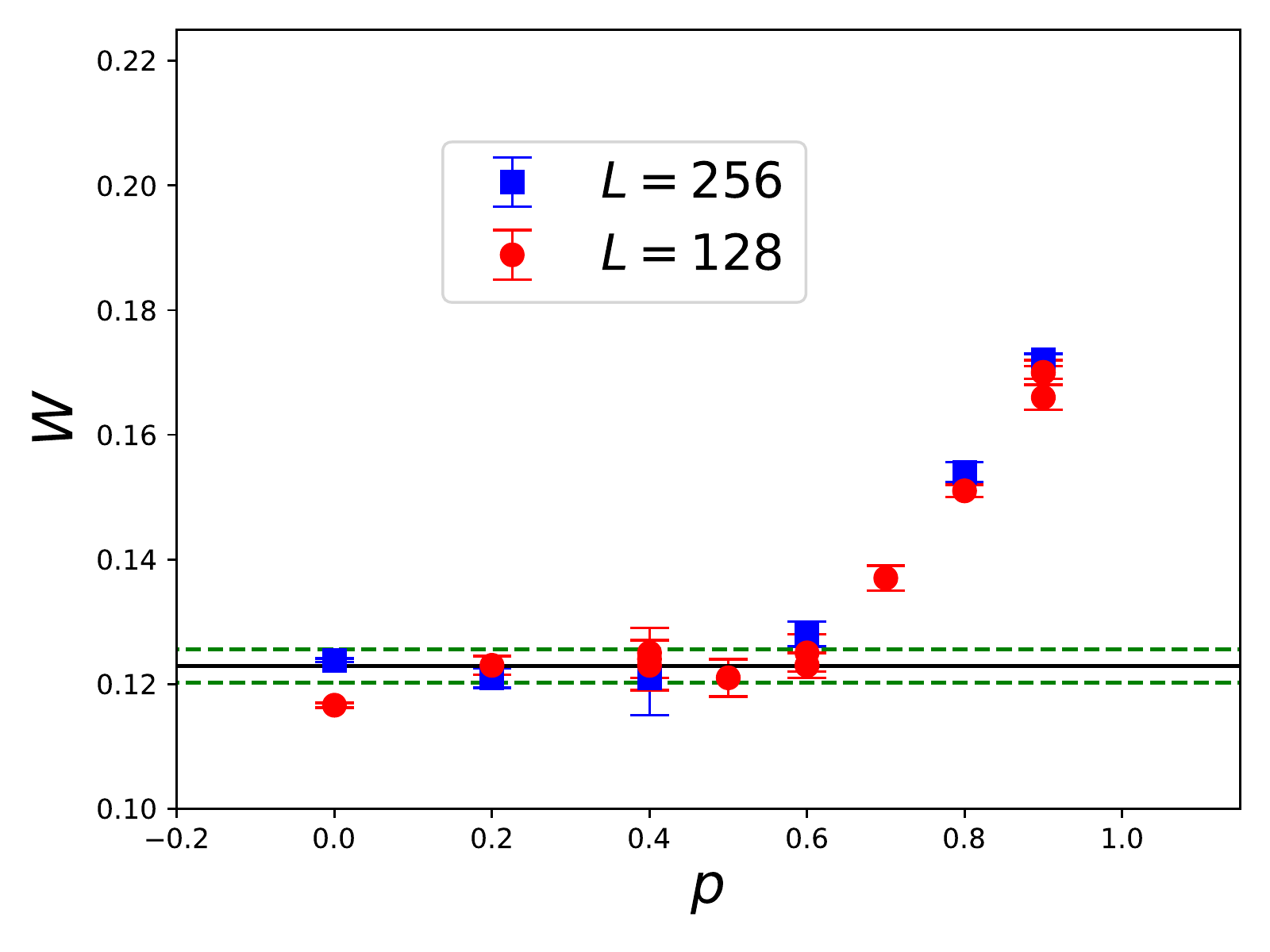}
\end{center}\vskip-0.5cm
\caption{$W$ as functions of $p$. The results are obtained from the analysis using the conventional uncorrelated $\chi^2$.
  The solid squares and solid circles are for $L=256$
  and $L=128$, respectively. For some values of $p$, the $L=128$ results contains
  those corresponding to $g_c$, the lower and upper bounds of $g_c$. The solid and dashed lines in the figure
  represent the mean and standard deviation for all the $W$ such that
  their corresponding $p$ satisfy $p < 0.7$ (including both those of $L=128$ and $L=256$). 
 }
\label{cfigW}
\end{figure}

In this study, we calculate the Wilson ratio $W$ of a 2D spin-1/2 antiferromagnetic Heisenberg model
with a specific quenched disorder, using the first principle nonperturbative quantum Monte Carlo
simulations. The employed disorder distribution has a tunable parameter $p$ which can be considered as a measure of
randomness. The $W$ of the clean case as well as that of $p=0.2,0.4,0.5,0.6,0.7,0.8,0.9$ are determined with
high precision. The critical dynamic exponents $z$ and the ground state energy densities $E_0$ are obtained as well.

Remarkably, for the considered system with the employed quenched disorder, the $p$ dependence of $W$ and $z$
seems to be complementary to each other. The obtained $z$ are likely to take a universal value for $p \ge 0.4$.
This agrees with the outcomes determined in Ref.~\cite{Skn04}. In addition, the calculated $W$ for $0 < p < 0.7 $
also have a trend of stay close to the result $W \sim 0.1243$ of the clean model ($p=0$). In particular,
the value of $W$ begins to increase sharply when $p$ is approaching $p=0.9$ where the Harris criterion is fulfilled.
Considering the fact that with what conditions the Harris criterion is valid is still not known \cite{Har74,Cha86,Mot00,San02d,Vaj02,Skn04,Yu05,San06d,Voj10,Yao10,Voj13,Voj14},
the results presented here may shed some light on setting up some
useful guidelines to decide whether the Harris criterion is valid for a given disorder distribution.

Apart from the subtlety of calculating $z$ described previously, the determination of $W$ is
extremely non-trivial as well. Indeed, the $W$ estimated here is based on Eq.~\ref{wilson_ratio} which contains
two constants $a$ and $b$. Since $a$ is a sub-leading coefficient in the
associated ansatz, it is sensitive to the range of $\beta$ used for the fits as well.
Careful strategy and resampling procedure are conducted in this study in order to calculate
$W$ accurately.

If the correlations among the data of various values of $\beta$ are ignored, then
the same resampling steps as well as the criterion of $\chi^2/{\text{DOF}} <3 $ (here the $\chi^2$
is the conventional uncorrelated $\chi^2$, not the correlated $\chi^2$ described in the main text) introduced in previous sections will lead to
figs.~\ref{cfigE_0},~\ref{cfigz}, and \ref{cfigW}. Remarkably, while the outcomes of $z$ shown in
fig.~\ref{cfigz} are slightly different from those in fig.~\ref{figz}, the $E_0$ and $W$
presented in figs.~\ref{cfigE_0} and \ref{cfigW} agree very well with the ones demonstrated
in figs.~\ref{figE_0} and \ref{figW}. In particular, the trend claimed from the analysis associated with the correlated $\chi^2$ regarding the $p$ dependence
of $z$ and $W$, namely being complementary to each other, 
is valid as well for the outcomes obtained using the conventional uncorrelated
$\chi^2$ (i.e. figs.~\ref{cfigz} and \ref{cfigW}). This observation seems to reconfirm the conclusions resulting from investigating
some lattice quantum chromodynamics data outlined in Refs.~\cite{Mic94,Mic95}.

To summarize, the outcomes resulting from the investigations carried out here,
specially the obtained numerical results of $E_0$, $z$, and $W$, are not only
important accomplishments, but also can be considered as benchmarks for future
related studies.

\vskip0.5cm
This study is partially supported by MOST of Taiwan.

~
~
~

~
~
~
~
~
~
~
~
~
~
~
~
~

\end{document}